\documentclass[11pt]{article}
\usepackage{graphicx}  
\usepackage{wrapfig}
\usepackage{epstopdf}
\newcommand{\GeV}{\mathrm{GeV}}
\newcommand{\MeV}{\mathrm{MeV}}
\newcommand{\pt}{p_{\rm{T}}}

\def\beq{\begin{equation}}
\def\eeq#1{\label{#1}\end{equation}}
\def\eeqn{\end{equation}}

\def\beqa{\begin{eqnarray}}
\def\eeqa#1{\label{#1}\end{eqnarray}}
\def\eeqan{\end{eqnarray}}

\let\bar=\overbar

\def\Dslash{\not{\hbox{\kern-4pt $D$}}}
\def\dslash{\not{\hbox{\kern-2pt $\del$}}}

\def\msb{{\bar{\ssstyle M \kern -1pt S}}}

\def\Title#1{\begin{center} {\Large {\bf #1} } \end{center}}

\begin{document}

\Title{Particle Identification with the ALICE detector at the LHC}

\bigskip


\begin{raggedright}  

{\it Chiara Zampolli\index{Zampolli, C.}  ALICE Collaboration\\
Istituto Nazionale di Fisica Nucleare, Sez. di Bologna\\
I-40126, ITALY}
\bigskip\bigskip
\end{raggedright}

\vspace{-1cm}

\section{Introduction}
The extreme conditions of high energy density and high temperature achieved in
Pb--Pb collisions at LHC energies are expected to realize a deconfined plasma of quarks and gluons 
(the so-called Quark-Gluon Plasma, QGP~\cite{QGP}), from which a phase transition to ordinary colourless hadronic matter takes place as a consequence of subsequent 
expansion and cooling down. ALICE (A Large Ion Collider Experiment, \cite{ALICE}) is the LHC experiment 
dedicated to the investigation of  the nature and
the properties of the QGP using heavy-ion collisions.
Especially, it is designed and built to cope with the high
track density environment expected in Pb--Pb collisions.
ALICE also can provide unique information on  low-$\pt$ pp physics
(thanks to the low material budget and low magnetic field of 0.5 T),
which makes the experiment complementary to CMS and ATLAS. 

ALICE addresses many observables,
spanning from the global characteristics of the events (such as multiplicity densities and rapidity distributions), to more 
specific QGP signals (like direct photons, charmonium and bottomonium). One of the basic requirement in order to carry out such measurements is an excellent particle identification (PID) performance. 
Making use of all known PID techniques, the ALICE detector is capable to identify hadrons and leptons over a very wide momentum range covering three
orders of magnitudes, from $\sim 100$ MeV/$c$ to $\sim 100$ GeV/$c$.

In the following sections, the ALICE detector will be briefly described (see Sec.~\ref{sec:ALICE}). The PID techniques used by the experiment will be presented in Sec.~\ref{sec:PID}.
Finally, a few examples of PID applications in physics analyses will be presented (Sec.~\ref{sec:Physics}). The conclusions 
will be drawn in Section~\ref{sec:Conclusions}.

\section{The ALICE detector}
\label{sec:ALICE}
\begin{figure}[t!]
\begin{center}
\begin{tabular}{cc}
\includegraphics[height=5cm]{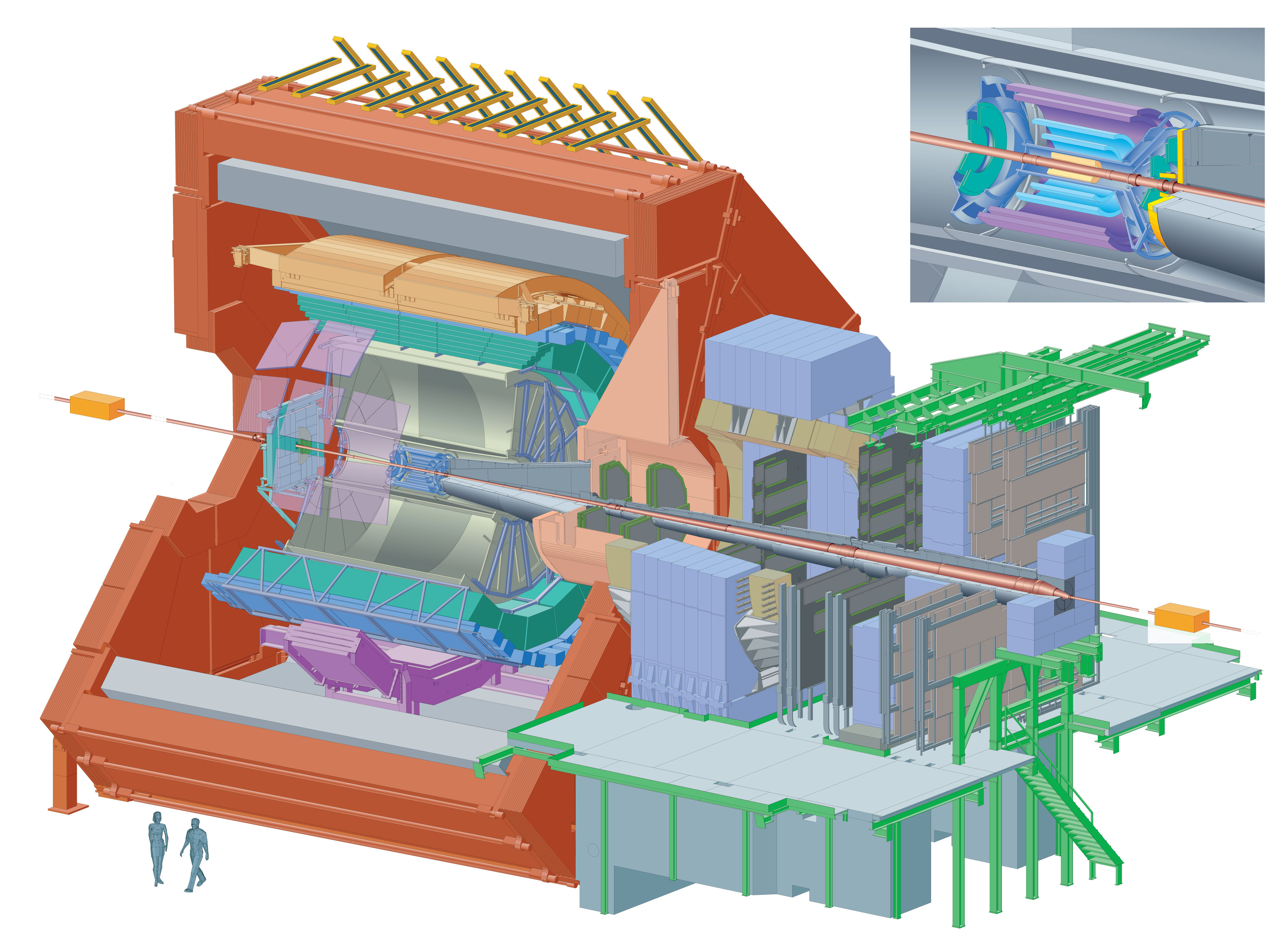} & \hspace{0.5cm}
\includegraphics[height=5cm]{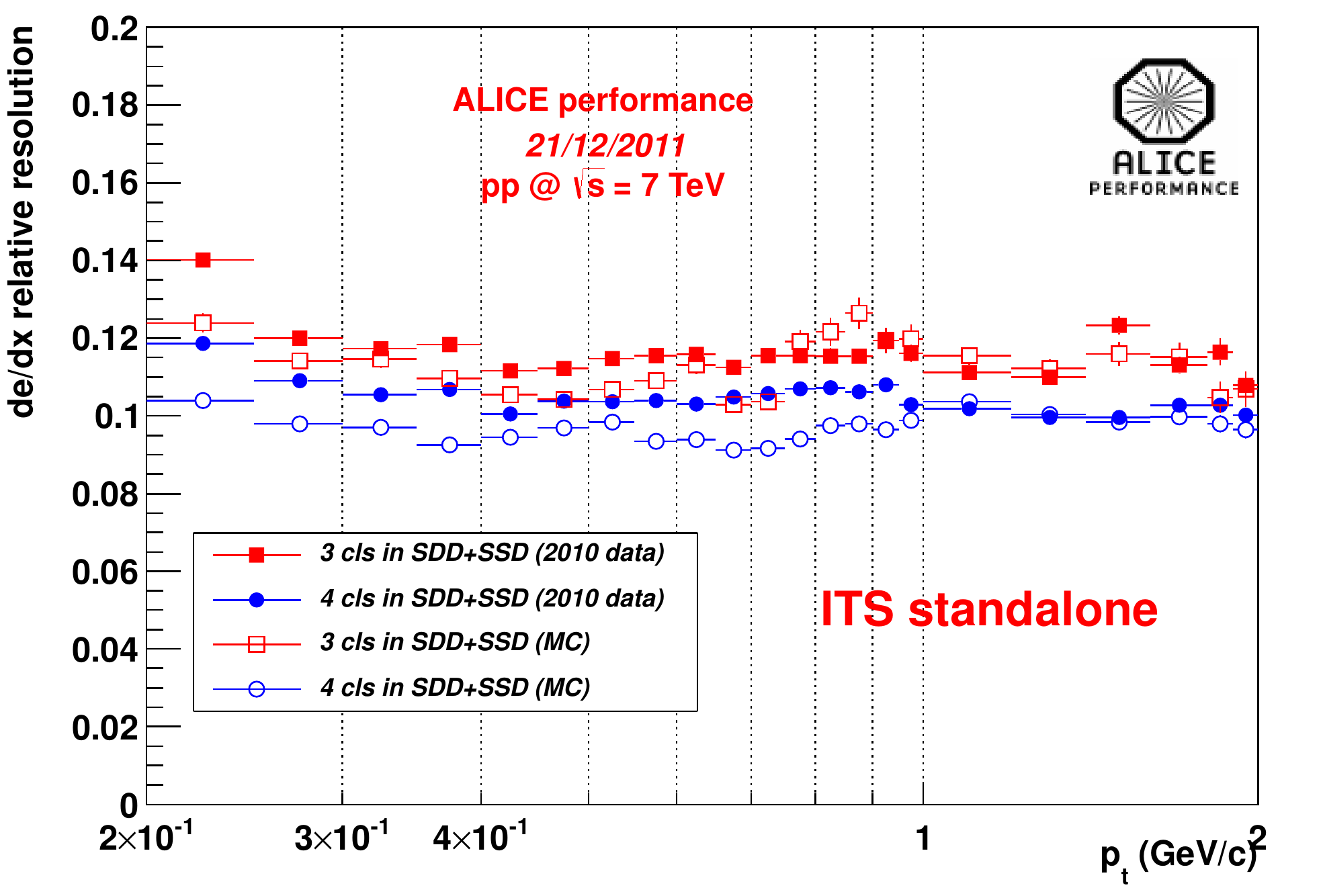}
\end{tabular}
\caption{Left panel: schematic view of the ALICE detector. Right panel: ITS $dE/dx$ resolution as a function of $\pt$ from data (filled markers) and Monte Carlo simulations (hollow markers). The performance is shown for the two different configurations, when either only three layers give a signal (red), or all of them (blue).}
\label{fig:ALICE}
\end{center}
\end{figure}

The left panel of Figure~\ref{fig:ALICE} shows a schematic view of the ALICE experiment. 

The detector closest to the beam pipe is the Inner Tracking System (ITS, shown also in the small inset), which is in charge of the reconstruction 
of the primary and secondary vertices. 
Three types of Si sensors are used for the ITS. 
The two innermost layers, for which a high granularity was needed 
to cope with the
requirement for the position resolution of primary and secondary
vertices,
consist of Silicon Pixel Detectors (SPD). They are followed by a pair of Silicon Drift Detectors (SDD), characterized by a very good multitrack reconstruction capability. 
Finally, two layers of Silicon 
Strip Detectors (SSD) complete the ALICE ITS. In addition to participate to the ALICE global tracking, the ITS is capable to perform standalone reconstruction with the advantage to recuperate the 
tracks lost 
in the global tracking due to the spacial
acceptance and the intrinsic $\pt$ cutoff of the outer detectors, and to
particle decay.

The ALICE Time Projection Chamber (TPC), following the ITS in radial direction, is the main ALICE tracking detector. Its tracking efficiency reaches $\sim 80\%$ in $|\eta| < 0.8$ with a momentum resolution $\sigma(\pt)/\pt \sim 5\%$, which gets down to $\sim 2.5\%$ up to $\pt = 10$ $\GeV/c$ (and increasing at higher transverse momenta) when combined with the ITS. 
Each track in
the TPC is  reconstructed using up to
a maximum of 159 space points, 
with a resolution of 0.8 mm in the xy plane, and 1.2 mm in the z direction. 

After TPC comes the Transition Radiation Detector (TRD), mainly dedicated to the electron identification. The TRD is followed by the Time Of Flight detector (TOF), 
at a radius of 3.7 m from the interaction point.

In the central $\eta$ region, ALICE has several
detectors, referred to as single-arm detectors, which have a limited
acceptance.
Namely, they are a Cherenkov RICH detector (the HMPID), a homogeneous 
photon spectrometer (PHOS), 
and a sampling electromagnetic calorimeter (EMCAL). At forward rapidities, a Photon Multiplicity Detector (PMD) and a muon spectrometer (MUON) are placed.

Some more detectors complete the ALICE setup, but they won't be described in these proceedings since they do not contribute to the particle identification of the experiment. For more details about them
and about the other ALICE detectors in general, see~\cite{ALICE}. 

\section{ALICE PID}
\label{sec:PID}
Out of the 16 detectors in ALICE, 6 provide particle identification information, using all the PID techniques known nowadays, implementing them at their state of art. Below, the 
different ALICE PID procedure will be presented. 
It is worth to mention that on top of PID technologies, ALICE identifies also cascades, V0 and kinks 
thanks to its excellent capability for tracking and
secondary vertex determination.

\subsection{PID in the central barrel}
\label{sec:CentralBarrel}
The central barrel detectors (i.e. those with full $\phi$ coverage) perform particle identification using the specific energy loss of a charged particle traversing a medium, the transition radiation emitted by charged particles when crossing
the boundary between two materials, and the time of flight that it takes to a charged particle to reach a detector's sensitive volume from the interaction point. 

$dE/dx$ measurements are provided by the last four layers of the ITS detector, i.e. the SDD and the SSD, thanks to their analog readout. A truncated mean is applied to the measurements, 
that is, an average
of the lowest two is taken if all the four layers gave a signal, or a weighted
average is taken
if only three are available. The ITS PID is performed in the low $\pt$ region, up to $\sim 1$ $\GeV/c$, and  
pions reconstructed in standalone mode can be identified down to $\sim 100$ $\MeV/c$. The right panel of Fig.~\ref{fig:ALICE} shows the $dE/dx$ resolution achieved by the ITS detector which stays around 10-15\% over the whole $\pt$ range.

\begin{figure}[t!]
\begin{center}
\begin{tabular}{cc}
\includegraphics[height=5cm]{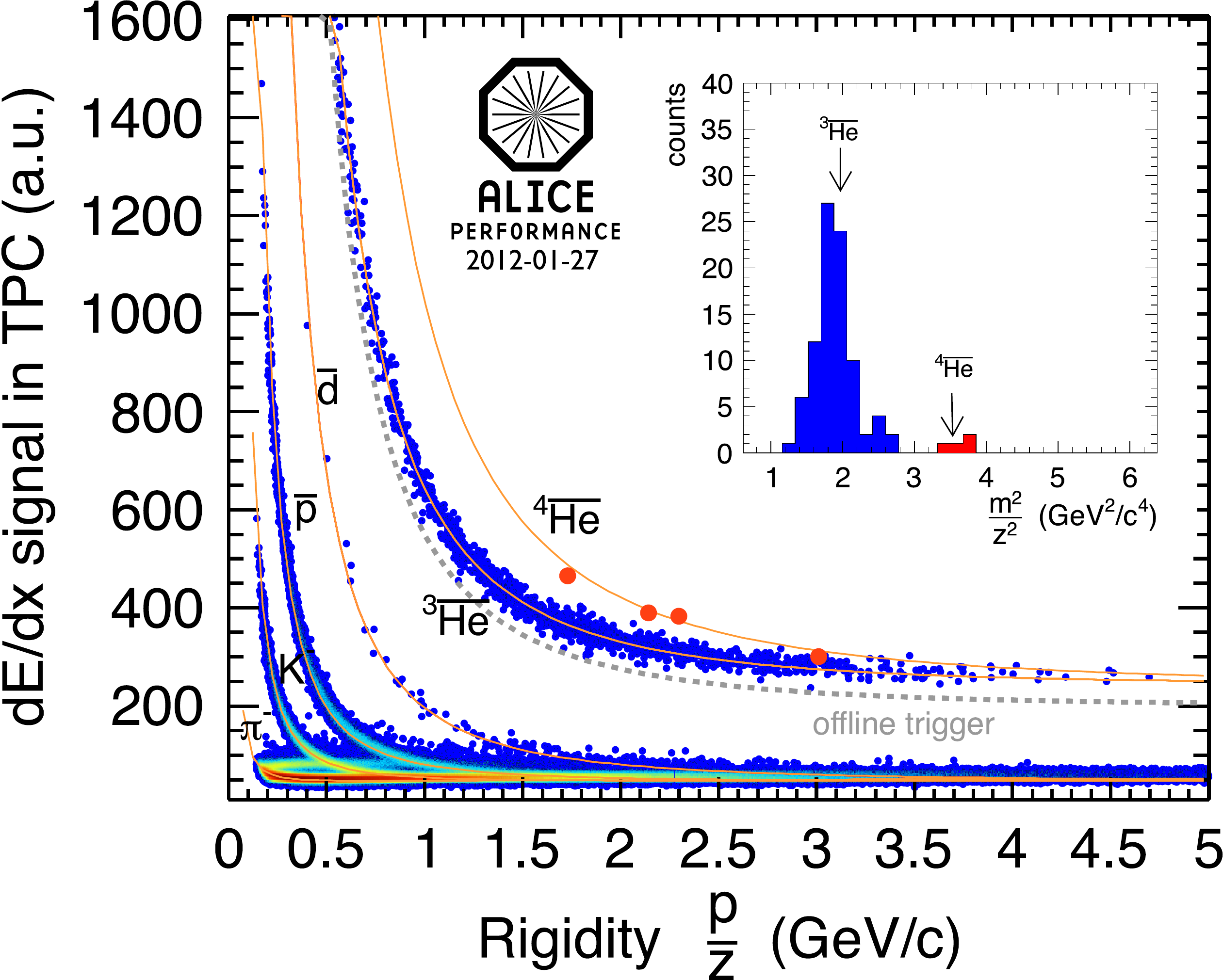} & \hspace{0.cm}
\includegraphics[height=5cm]{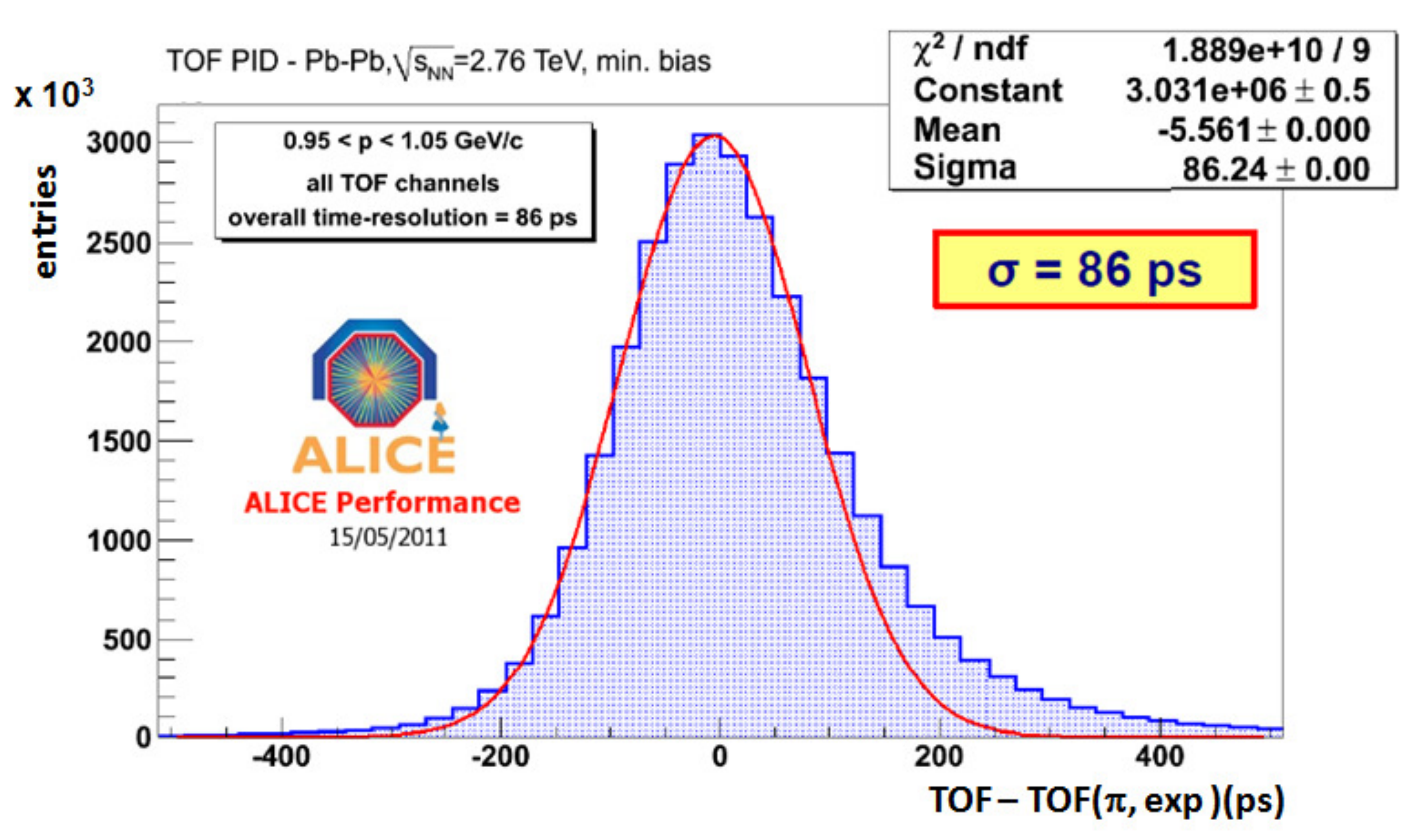}
\end{tabular}
\caption{Left panel: Energy loss measured by the TPC as a function of rigidity for negative tracks. The inset shows the distribution of $m^2/z^2$ for the light nuclei candidates obtained with the TOF detector. Right panel: Difference between the measured and the expected time of flight in the pion hypothesis. Superimposed, the gaussian fit of the distribution.}
\label{fig:TPC}
\end{center}
\end{figure}

The ALICE TPC detector adds PID information using specific energy loss measurements as well. Also in this case, a truncated mean is applied over the maximum number of 159 cluster information. The performance is excellent, with a resolution of $\sim 5\%$ calculated for isolated tracks, in the cases when 159 space points were available. In addition to the identification of charged hadrons up 
to $\pt \sim$ 1 -- 2 $\GeV/c$, the TPC wide dynamic range (up to 26 MIP) allows to identify light nuclei, as shown in the left panel of Fig.~\ref{fig:TPC}. Moreover, while in the $1/\beta^2$ Bethe-Bloch region 
of the $dE/dx$ distribution particle
identification for individual tracks is possible, in the region of
relativistic rise
a statistical approach is utilized, allowing the TPC to identify charged hadrons up to $\pt$ of a 
few tens of $\GeV/c$.

Electron identification in ALICE is carried out by the TRD in the momentum region $p > 1$ $\GeV/c$, with a pion rejection factor of $~100$. The PID relies on a 1-dimentional
likelihood approach, which makes it possible to distinguish between pions and electrons due to the different shapes of the signals they release in the detector\footnote{The signal from electrons
in a TRD detector is characterized by a further peak at late times due to the presence of transition radiation photons, which is absent in the case of pions.}.

Charged hadrons in the intermediate momentum range (i.e. up to a few $\GeV/c$) are identified in ALICE by the TOF detector. In this case, the mass (and as a consequence the identity) of 
a particle is obtained by combining the measurement of its time of flight (from TOF) and its momentum (from ITS and TPC). The reference time of the event is given by a combination of the 
the event time information from the ALICE T0 detector, and the one 
estimated from the particle arrival times measured by TOF.
The right panel of Fig.~\ref{fig:TPC} shows the TOF 
resolution for the identification of pions (the most abundant particle specie) in terms of the difference between the measured time of flight, and the expected one 
calculated from the  track length and
momentum assuming that the particle is a pion.
As one can see, the detector performance is outstanding, with $<90$ ps of resolution. 

\subsection{PID with single-arm detectors}
\label{sec:Single-Arm}
\begin{figure}[t!]
\begin{center}
\begin{tabular}{cc}
\includegraphics[height=5cm]{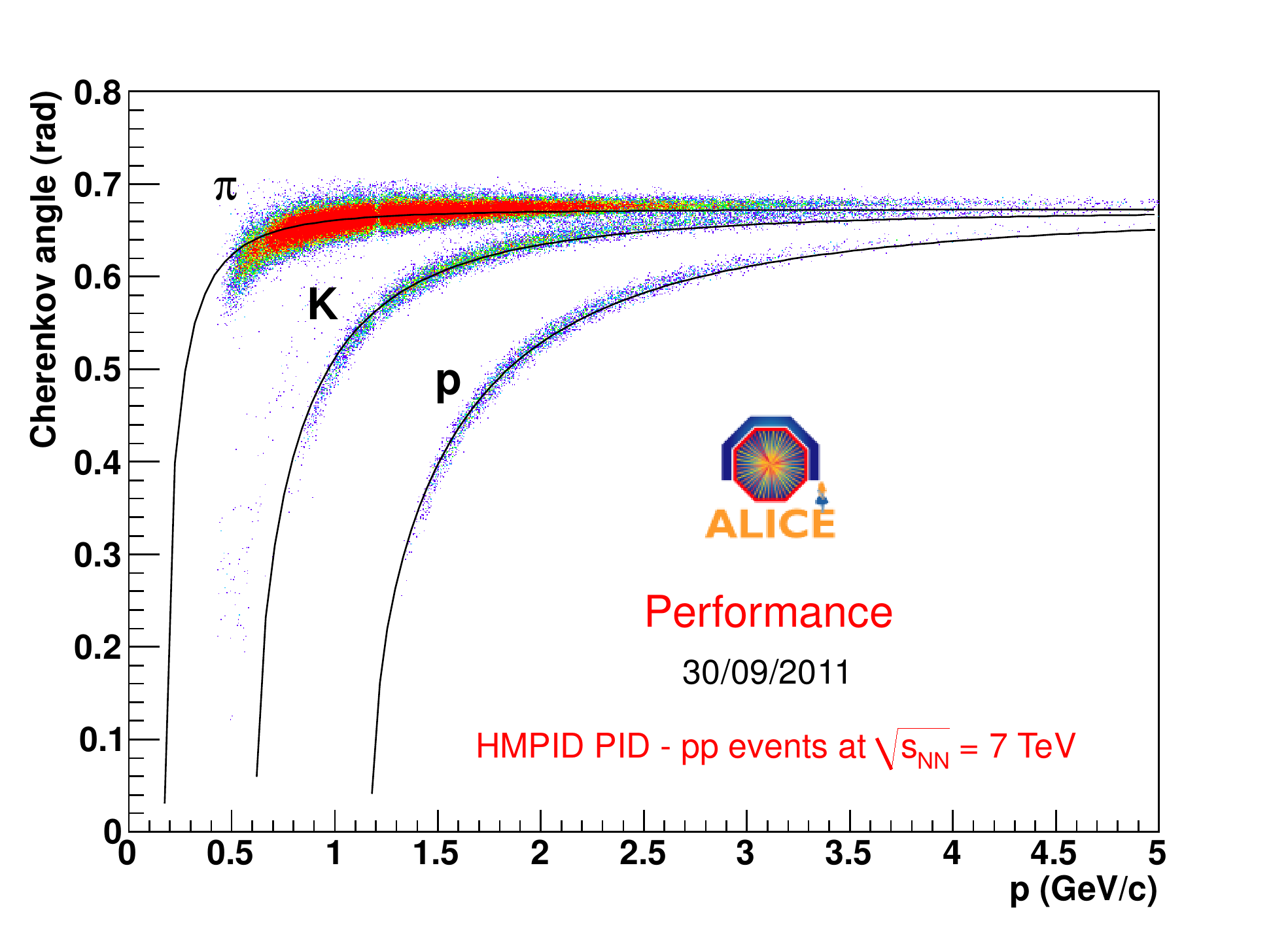} & \hspace{0.cm}
\includegraphics[height=5cm]{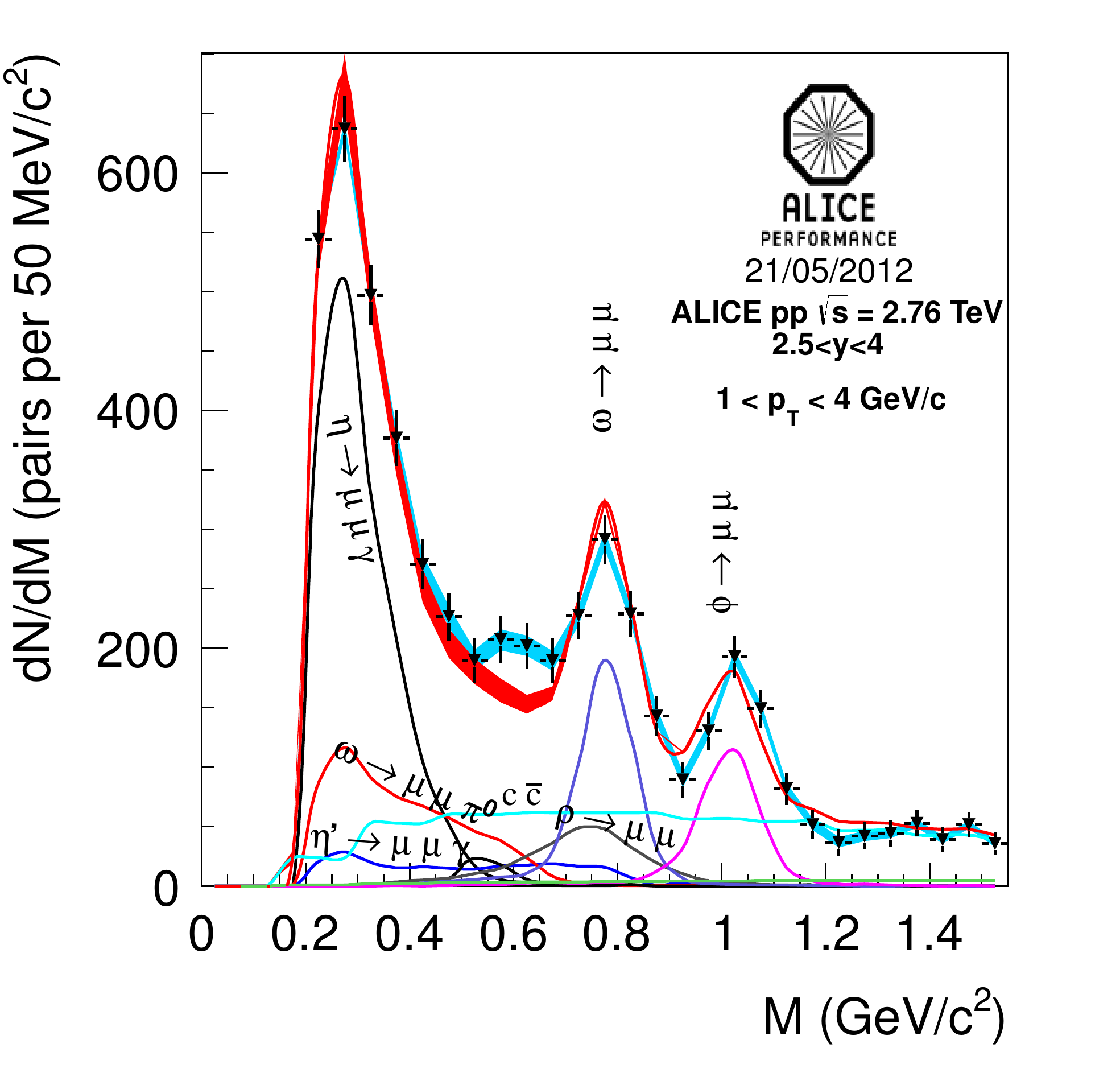}
\end{tabular}
\caption{Left panel: Cherenkov angle measured by the HMPID as a function of $p$. Right panel: Invariant mass distribution of muon pairs measured by the MUON detector.}
\label{fig:HMPID}
\end{center}
\end{figure}

Hadron identification at high momenta (up to 3--5 $\GeV/c$ depending on the specie) is performed by the  HMPID detector. This is a single-arm proximity focusing RICH, which 
determines the $\beta$ of a particle
from the measurement of the Cherenkov angle. This information is then combined with the momentum measured by the TPC and ITS to assign an identity to the particle. The left panel of 
Fig.~\ref{fig:HMPID} shows the Cherenkov angle measured by the HMPID as a function of $p$. As one can see, the $\pi$, K and p bands are clearly distinguishable.

The two ALICE electromagnetic calorimeters, PHOS and EMCAL, have partial $\eta$ and $\phi$ coverage as well. They measure $\gamma$ up to 100 and 250 $\GeV$ respectively. The EMCAL
is also used in ALICE to help hadron rejection when identifying electrons, thanks to the $E/p$ distribution characteristically peaked at 1 only for electrons due to their small mass.

At forward rapidities where the multiplicities are too high to use calorimetry, photons are identified in ALICE also using the PMD, through the pre-shower method.

On the opposite side of the experiment, a MUON spectrometer reconstructs and identifies muons in the momentum range $p>4$ $\GeV$. Hadron rejection is possible
requiring matching between the tracks reconstructed by the tracking chambers with one track segment in the triggering chambers. Moreover, geometrical and topological cuts are applied in order to reduce
contamination from fake tracks, and Monte Carlo simulations are used to estimate muon contributions from hadron decays. The right panel of Fig.~\ref{fig:HMPID} shows the invariant mass distribution of $\mu$
pairs ($1 < \pt < 4$ $\GeV/c$) reconstructed and identified by the MUON detector after background subtraction. The various contributions to the spectrum are shown.

\section{Physics Results with PID}
\label{sec:Physics}
Many of the ALICE physics results rely on Particle Identification. Since a comprehensive review is not possible in these proceedings, only a few examples will be presented. For more information
about the ALICE results, see~\cite{Abelev}.

Hadron identification is one of the key elements in the femtoscopy studies. These are aimed at the measurement of the size and shape of the emitting source especially important in Pb--Pb
collisions, 
by utilizing the Bose-Einstein correlations between
identical particles.
When pion pairs are used, the particle identification response of the TPC is used. Studies are also made for charged kaons, 
and in this case the analysis depends on both TPC and TOF PID. Recently, results using neutral kaons have also become available despite the smaller statistics, 
In this case, the topological PID is exploited for the K$_0$
identification.

\begin{figure}[t!]
\begin{center}
\begin{tabular}{cc}
\includegraphics[height=5cm]{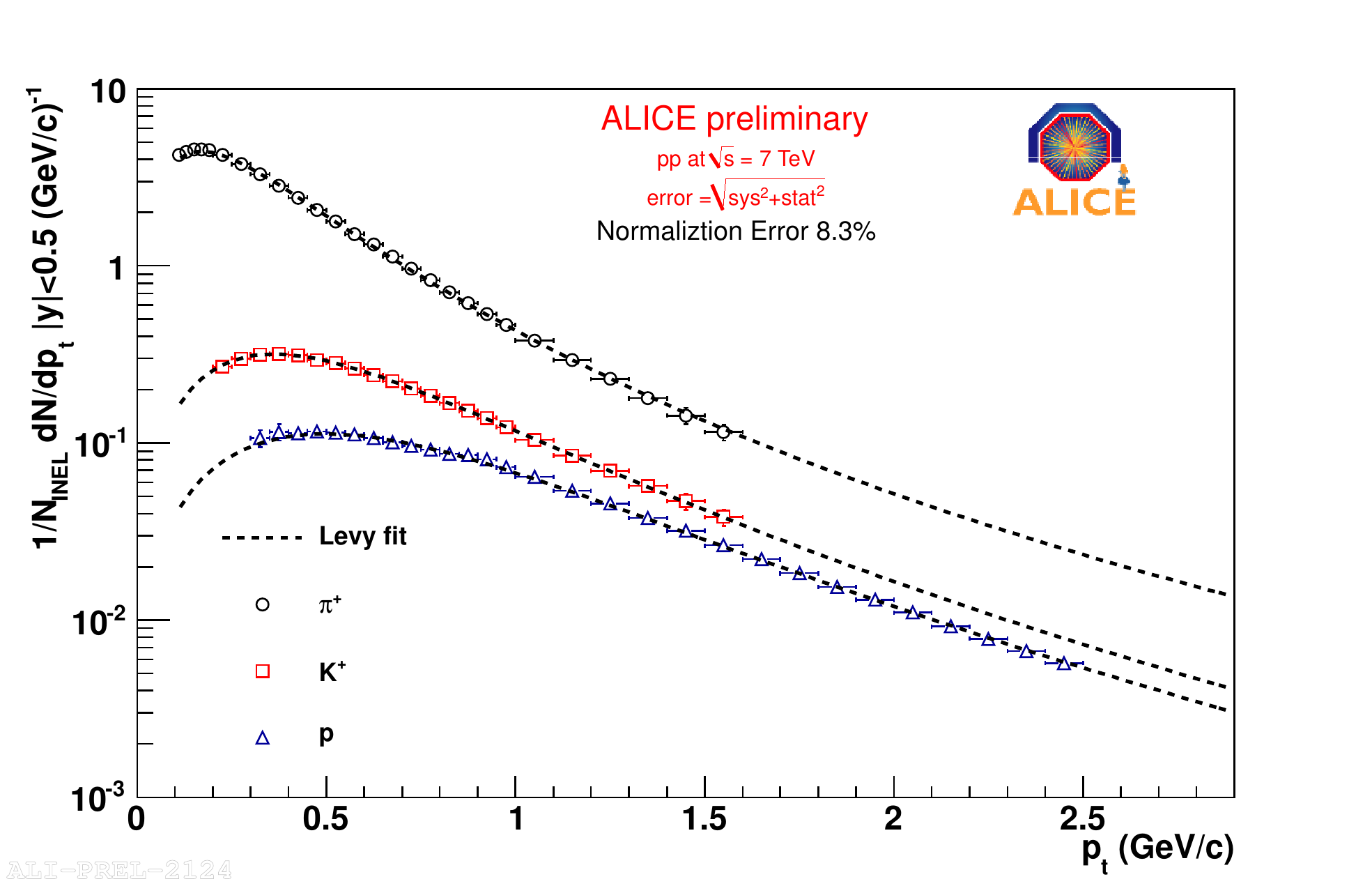} & \hspace{0.cm}
\includegraphics[height=5cm]{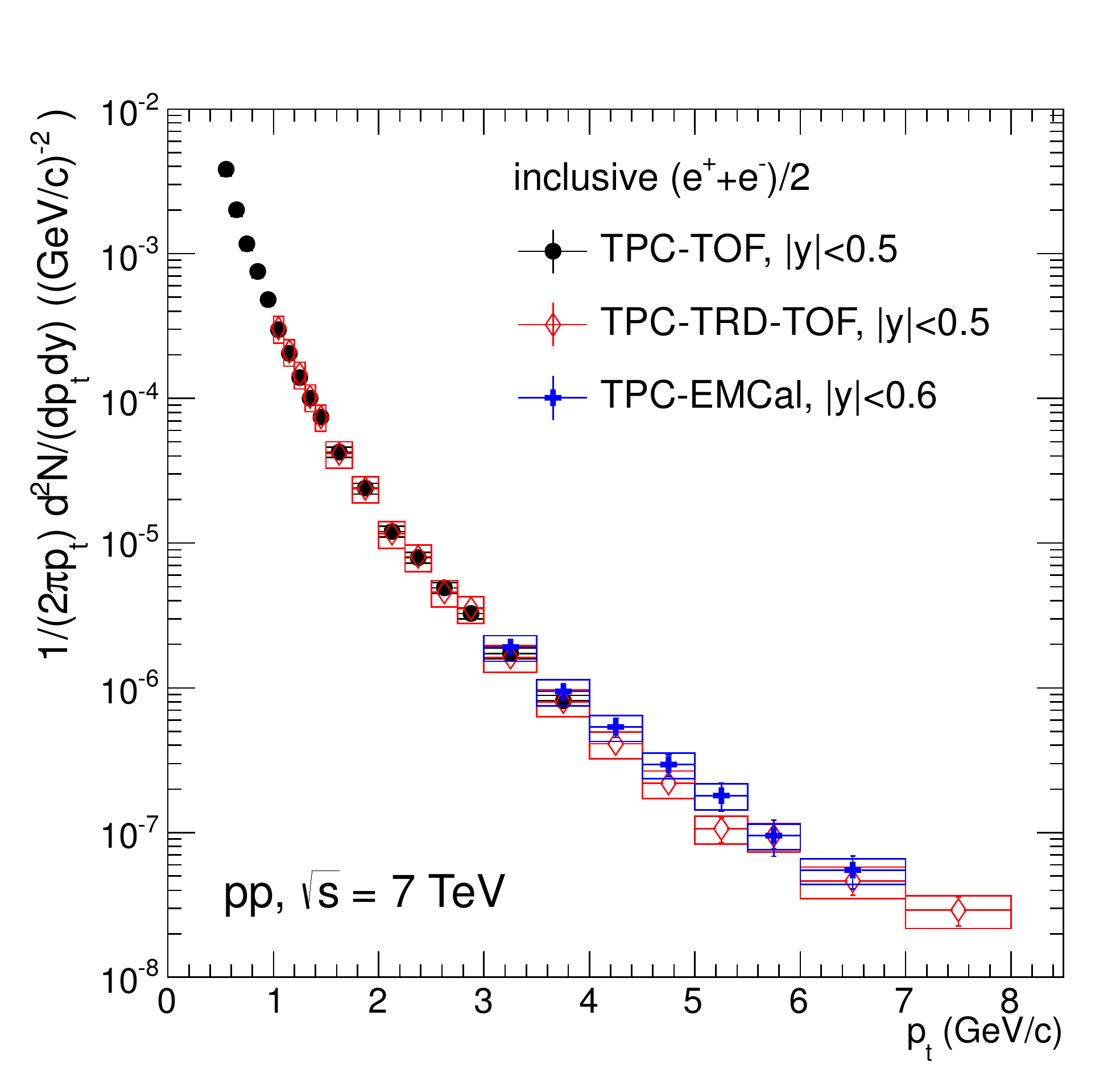}
\end{tabular}
\caption{Left panel: Identified pions, kaons and protons using the combination of the information from ITS, TPC and TOF. The L\'evy fit are superimposed. Right panel: Inclusive electron spectrum
measured by three ALICE analysis using different detectors.}
\label{fig:spectra}
\end{center}
\end{figure}

The knowledge of the particle composition of the low $\pt$ hadrons at mid-rapidity is important in order to understand the hadronization
mechanisms. ALICE combines the responses of its different PID detectors in order to build the identified charged hadrons spectra as shown in the left panel 
of Fig.~\ref{fig:spectra} for the case of positive pions, kaons and protons. 
Here, the ITS, TPC and TOF PID information\footnote{The inclusion in the analysis of the HMPID is ongoing.} are used allowing to extend the PID reach of the experiment. The fit with the L\'evy function 
are superimposed.

The right panel of Fig.~\ref{fig:spectra} shows the inclusive electron spectrum measured by ALICE in pp collisions at 7 TeV. Here, the results from three different analysis based on the use
of different combination of detectors are superimposed. Up to 3 $\GeV/c$, the TPC and TOF are used for the identification of electrons; in the range $1 < \pt < 8$ $\GeV/c$, the TRD is 
included in the analysis. Finally in the $\pt$ range between 3 and 7 $\GeV/c$ the EMCAL is used together with the TPC. 
From the figure one can see their
remarkable agreement in the overlapping ranges, and will infer the
complementarity of the ALICE PID techniques.

\vspace{-0.2cm}
\section{Conclusions}
\label{sec:Conclusions}
The ALICE experiment at the LHC is endowed with many different PID detectors. Thanks to the different momentum range that they cover, the ALICE
PID capability is extended to a broad momentum interval.
The use of the most up-to-date techniques and technologies makes ALICE PID capabilities unique. Many analysis
reckon on PID results, both in pp and Pb--Pb collisions. The outstanding PID results demonstrates that ALICE is fully complying with its wide and varied physics program.

\end{document}